\documentclass[10pt]{elsart}     
\usepackage{amssymb,graphicx,epsfig,pstricks,rotating}      
\newcommand{\gsim}{\stackrel{\scriptstyle >}{\phantom{}_{\sim}}}      
      
\newcommand\eqn[1]{(\ref{#1})}

\newcommand{\dslash}{{\partial\!\!\!/}}

\newcommand{\figurewidth}{\linewidth}     
   
\begin{document}     
\begin{frontmatter}     
\title{    
Modern compact star observations and the quark matter    
equation of state    
}     
\date{\today}     
     
\author[argonne,rostock,gsi]{T.~Kl\"ahn,}     
\author[rostock,wroclaw,dubna1]{D.~Blaschke,}     
\author[lulea]{F.~Sandin,}     
\author[tuebingen]{Ch.~Fuchs,}     
\author[tuebingen]{A.~Faessler,}     
\author[dubna2,yerevan]{H.~Grigorian,}     
\author[rostock]{G.~R\"opke,}     
\author[mpe]{J.~Tr\"umper}     
     
\address[argonne]{     
Physics Division,      
Argonne National Laboratory,     
Argonne, IL 60439-4843,      
USA}     
\address[rostock]{     
Institut f\"ur Physik,      
Universit\"at Rostock,     
D-18051 Rostock,      
Germany}     
\address[gsi]{     
Gesellschaft f\"ur Schwerionenforschung mbH (GSI), \\     
D-64291 Darmstadt,      
Germany}     
\address[wroclaw]{     
Institute for Theoretical Physics, University of Wroclaw \\     
PL-50-204 Wroclaw,      
Poland}     
\address[dubna1]{     
Bogoliubov  Laboratory of Theoretical Physics,\\     
Joint Institute for Nuclear Research,      
RU-141980 Dubna,      
Russia}     
\address[lulea]{Department of Physics,      
Lule{\aa}  University of Technology,      
SE-97187 Lule\aa,      
Sweden}     
\address[tuebingen]{     
Institut f\"ur Theoretische Physik,      
Universit\"at T\"ubingen,\\     
D-72076 T\"ubingen,      
Germany}     
\address[dubna2]{     
Laboratory for Information Technologies,\\     
Joint Institute for Nuclear Research,      
RU-141980 Dubna,      
Russia}     
\address[yerevan]{     
Department of Physics,     
Yerevan State University,     
375049 Yerevan,     
Armenia}     
\address[mpe]{  
Max-Planck-Institut f\"ur Extraterrestrische Physik,  
D-85741 Garching, Germany}  
\begin{abstract}      
We present a hybrid equation of state (EoS) for dense   
matter that satisfies phenomenological constraints   
from modern compact star (CS) observations which indicate   
high maximum masses ($M\sim 2 M_\odot$) and large radii  
($R> 12$ km).  
The corresponding isospin symmetric EoS  
is consistent with flow data analyses of heavy-ion collisions  
and a deconfinement transition at $\sim 0.55$ fm$^{-3}$.  
The quark matter phase is described by a 3-flavor   
Nambu--Jona-Lasinio model that accounts for scalar diquark  
condensation and vector meson interactions while  
the nuclear matter phase is obtained within the   
Dirac-Brueckner-Hartree-Fock (DBHF) approach using the Bonn-A   
potential.   
We demonstrate that  both pure neutron stars and   
neutron stars with quark matter cores   
are consistent with modern CS observations.  
Hybrid star configurations with a CFL quark core are unstable  
within the present model.   
\\[5mm]  
\noindent PACS number(s):  04.40.Dg, 12.38.Mh, 26.60.+c, 97.60.Jd     
\end{abstract}      
\end{frontmatter}     
\newpage     
   
\section{Introduction}   
Recently, new observational limits for the mass and the mass-radius   
relationship of compact stars have been obtained which provide  
stringent constraints on the equation of state of strongly interacting   
matter at high densities, see \cite{Klahn:2006ir} and references therein.  
In particular, the high mass of $M=2.1\pm 0.2~M_\odot$ for the pulsar  
J0751+1807 in a neutron star - white dwarf binary system \cite{NiSp05}    
and  the large radius of $R > 12$ km for the   
isolated neutron star RX J1856.5-3754 (shorthand: RX J1856)   
\cite{Trumper:2003we} point to a stiff equation of state   
at high densities.  
Measurements of high masses are also reported for compact stars in low-mass  
X-ray binaries (LMXBs) as, e.g.,  
$M=2.0\pm 0.1~M_\odot$ for the compact object in 4U 1636-536  
\cite{Barret:2005wd}.  
For another LMXB, EXO 0748-676,   
constraints for the mass $M\ge 2.10\pm 0.28~M_\odot$  
{\it and} the radius $R \ge 13.8 \pm 0.18$ km 
have been reported \cite{Ozel:2006km}.  
The status of these data is, however, unclear since the observation of a   
gravitational redshift $z=0.35$ in the X-ray burst spectra \cite{Cottam:2002}   
could not be confirmed thereafter despite numerous attempts \cite{Mendez:2006}.  
We exclude possible constraints from LMXBs from the discussion in the present   
paper as their status is not settled and they would not tighten the mass and   
mass-radius limits provided by J0751+1807 and RX J1856, respectively.  
It has been argued \cite{Trumper:2003we,Ozel:2006km} that deconfined quark   
matter cannot exist in the centers of compact stars with masses and radii as   
reported for these objects.   
In view of recent works on the quark matter EoS, however, this claim appears   
to be premature \cite{Alford:2006vz}.  
  
In the present paper, we demonstrate that   
the present-day knowledge of hydrodynamical  properties of dense matter   
allows to construct hybrid EsoS with a critical density of the   
deconfinement phase transition low enough to allow for extended quark matter   
cores and stiff enough to comply with the new mass measurements of  
compact stars.  
It has been shown recently by Alford et al. \cite{Alford:2004pf} that  hybrid  
stars can masquerade as neutron stars once the  
parameters of a generic phenomenological quark matter EoS have been chosen   
appropriately.   
While in \cite{Alford:2004pf} the APR EoS \cite{APR} for the hadronic   
phase has been used, we base our investigation on a nuclear EoS obtained    
from the ab initio DBHF approach using the Bonn A potential   
\cite{DaFuFae04,DaFuFae05}  
which results in star configurations with larger radii and masses.    
{The DBHF EoS is soft at moderate densities (compressibility K=230 MeV)   
\cite{DaFuFae04,boelting99} but tends to become stiffer at high   
densities. At densities up to 2-3 times nuclear saturation   
density it is in agreement with constraints   
from heavy ion collisions based on collective flow   
\cite{Danielewicz:2002pu,stoicea04}   
and kaon production \cite{kaons}. However, at higher densities this EoS   
seems to be too repulsive concernig the flow constraint.}  
As we will show in the present paper, the problems of this EoS with  
an early onset of the nuclear direct Urca (DU) process and a violation of the  
flow constraint for heavy-ion collisions at high densities can be solved by  
adopting a phase transition to quark matter.  
    
We have no first principle information from QCD about the quark matter EoS  
in the nonperturbative domain close to the chiral/ deconfinement transition  
at zero temperature and finite density which would be required for   
an ab-initio study of the problem whether deconfined quark matter can exist   
in neutron stars or not.  
Therefore it is desireable to develop microscopic approaches to the  
quark matter EoS on the basis of effective models implementing, as far as  
possible, QCD symmetries into the model Lagrangian.   
Such an approach should in particular be capable of describing the   
phenomena of dynamical chiral symmetry breaking/restoration and color   
superconductivity in dense quark matter \cite{Blaschke:2006xt},  
going thus beyond the traditional   
bag model approaches widely used for phenomenological studies of quark matter  
in compact stars, see \cite{Glendenning:1997wn,Weber:1999qn} and references therein.   
A powerful modeling of QCD in the perturbative and nonperturbative  
domain is obtained using Dyson-Schwinger equations   
\cite{Roberts:1994dr,Alkofer:2000wg} which recently have been extended also to   
the finite density domain \cite{Blaschke:1997bj,Roberts:2000aa,Nickel:2006kc}.   
They appear as most promising candidate theories for a future systematic QCD   
modeling of dense QCD matter in compact stars.  
  
For the present study, however, we will employ  
the Lagrangian of a Nambu--Jona-Lasinio (NJL) model with chiral   
symmetry which is dynamically broken in the nonperturbative vacuum and   
restored at finite temperatures in accordance with lattice QCD simulations.  
Therefore, the application of the NJL model to finite densities where  
presently no reliable lattice QCD simulations exist, can be regarded as   
a step beyond the bag model phenomenology towards a dynamical approach  
to nonperturbative quark matter effects in compact stars, see   
\cite{Buballa:2003qv}  
and references therein.   
Although the standard NJL model is quite simple,  
based on a few parameters, such as the momentum cutoff and coupling constants   
for a set of interaction channels, it offers possibilities for generalizations,  
like a density-dependence of the cutoff parameter \cite{Baldo:2006bt},   
momentum-dependent formfactors \cite{Schmidt:1994di} and an infrared cutoff  
to account for confinement  \cite{Blaschke:1998ws,Lawley:2006ps}.  
  
In this work, we will use a standard three-flavor NJL model, where in   
contrast to the approach used in Ref. \cite{Alford:2004pf} we 
base our investigation on selfconsistently determined quark masses  
and pairing gaps \cite{Blaschke:2005uj}, similar to the parallel developments   
in Refs. \cite{Ruster:2005jc,Abuki:2005ms,Warringa:2005jh}.  
This approach has the advantage that it allows, e.g., to distinguish two- and   
three-flavor phases in quark matter   
(for a first discussion, see \cite{Gocke:2001})  
and conclusions about the presence of gapless phases at zero   
temperature as a function of the coupling strengths in the   
current-current-type interaction of the model Lagrangian \cite{Sandin:2005um}.  
Moreover, we will investigate the question of the stability of neutron stars   
with a color superconducting quark matter core in the celebrated CFL phase,   
for which a number of phenomenological applications have been studied, in   
particular the cooling problem   
\cite{Blaschke:1999qx,Page:2000wt,Blaschke:2000dy,Alford:2004zr},  
gamma-ray bursts \cite{Berdermann:2005yn,Drago:2005qb},  
and superbursts \cite{Page:2005ky}.  
We will confirm in this work the earlier result by Baldo et al.  
\cite{Baldo:2002ju} and Buballa et al. \cite{Buballa:2003et} that  
a CFL quark matter core renders the hybrid star unstable.  
  
Here we generalize the model \cite{Blaschke:2005uj} by including an isoscalar  
vector meson current which, similar to the Walecka model for nuclear matter,   
leads to a stiffening of the quark matter EoS. Increasing the scalar diquark   
coupling constant leads to a lowering of the phase transition density.   
It is the aim of the present work to determine the unknown coupling strengths  
in both these channels such that an optimal hybrid star EoS is obtained.  
It fulfills all recently developed constraints from modern compact star   
observations  \cite{Klahn:2006ir,Ozel:2006km} while providing   
sufficient softness of the isospin-symmetric limit of this EoS as required   
from the analysis of  heavy-ion collision   
transverse flow data \cite{Danielewicz:2002pu,stoicea04} and   
$K^+$ production data \cite{kaons}.  
    
\section{Equation of state}   
\label{sec:eos}   
   
The thermodynamics of the deconfined quark matter phase is described   
within a three-flavor quark model of Nambu--Jona-Lasinio (NJL) type.   
The path-integral representation of the partition function is given by  
\begin{eqnarray}   
\label{Z}  
Z(T,\hat{\mu})&=&\int {\mathcal D}\bar{q}{\mathcal D}q   
\exp \left\{\int_0^\beta d\tau\int d^3x\,\left[   
        \bar{q}\left(i\/\dslash-\hat{m}+\hat{\mu}\gamma^0\right)q+  
{\mathcal L}_{\rm int}   
\right]\right\}, \label{Zqqbar} \\   
{\mathcal L}_{\rm int} &=& G_S\Bigg[   
        \sum_{a=0}^8(\bar{q}\tau_aq)^2 +   
        \eta_V(\bar{q}i\gamma^0q)^2 \nonumber \\  
        &&+\eta_D\!\!\!\!\sum_{A=2,5,7}\!\!\!   
        (\bar{q}i\gamma_5\tau_A\lambda_AC\bar{q}^T)   
        (q^TiC\gamma_5\tau_A\lambda_Aq),   
\Bigg],   
\end{eqnarray}   
where $\hat{\mu}$ and $\hat{m}={\rm diag}_f(m_u,m_d,m_s)$ are the   
diagonal chemical potential and current quark mass matrices.   
For $a=0$, $\tau_0=(2/3)^{1/2}{\mathbf 1}_f$, otherwise $\tau_a$ and   
$\lambda_a$  are Gell-Mann matrices acting in flavor  
and color spaces, respectively.   
$C=i\gamma^2\gamma^0$ is the charge conjugation operator and   
$\bar{q}=q^\dagger\gamma^0$. $G_S$, $\eta_V$, and $\eta_D$ determine   
the coupling strengths of the interactions.   
   
The interaction terms represent current-current interactions in the color   
singlet scalar and vector meson channels, and the scalar color and flavor   
antitriplet diquark channel.   
In the choice of the four-fermion interaction channels we have  
omitted the chiral partner terms which would be necessary to establish  
the chiral symmetry of the Lagrangian as their contributions  
to the thermodynamical potential vanish at the mean-field (Hartree)  
level to which we restrict ourselves in the present paper.  
The model is similar to the models in   
\cite{Blaschke:2005uj,Ruster:2005jc,Abuki:2005ms}, except that we   
include also the isoscalar vector meson channel.   
We follow the argument given in \cite{Blaschke:2005uj}, that the  
U$_A$(1) symmetry breaking in the pseudoscalar isoscalar meson sector  
is dominated by quantum fluctuations and no 't~Hooft determinant  
interaction needs to be adopted for its realization.  
   
After bosonization using Hubbard-Stratonovich   
transformations, we obtain an exact transformation of the original   
partition function \eqn{Zqqbar}.   
The transformed expression constitutes the   
starting point for powerful approximations, defined as truncations of the   
Taylor expanded action functional to different orders in the   
collective boson fields.    
In the following, we use the mean-field (MF) approximation. This means   
that the bosonic functional integrals are omitted and the collective fields   
are fixed at the extremum of the action. The corresponding mean-field   
thermodynamic potential, from which all thermodynamic quantities can   
be derived, is given by   
\begin{eqnarray}   
\Omega_{MF}(T,\mu) &=&   
        -\frac{1}{\beta V}\ln Z_{MF}(T,\mu)\nonumber\\   
&=& \frac{1}{8 G_S}\left[\sum_{i=u,d,s}(m^*_i-m_i)^2    
  - \frac{2}{\eta_V}(2\omega_0^2+\phi_0^2)  
        +\frac{2}{\eta_D}\sum_{A=2,5,7}|\Delta_{AA}|^2\right]    
        \nonumber \\   
&&-\int\frac{d^3p}{(2\pi)^3}\sum_{a=1}^{18}   
        \left[E_a+2T\ln\left(1+e^{-E_a/T}\right)\right]   
        + \Omega_l - \Omega_0~.   
\label{potential}   
\end{eqnarray}   
Here, $\Omega_l$ is the thermodynamic potential for electrons and muons,   
and $\Omega_0$ is a divergent term that is subtracted in order to get zero   
pressure and energy density in vacuum ($T=\mu=0$). The quasiparticle   
dispersion relations, $E_a(p)$, are the eigenvalues of the Hermitean matrix   
\begin{equation}   
{\mathcal M} = \left[   
\setlength\arraycolsep{-0.01cm}   
\begin{array}{cc}   
        -\vec{\gamma}\cdot\vec{p}- \hat{m}^*+\gamma^0\hat{\mu}^* &    
        \gamma_5 \tau_A \lambda_A \Delta_{AA} \\   
        - \gamma_5 \tau_A \lambda_A\Delta_{AA}^* &    
        -\vec{\gamma}^T\cdot\vec{p}+\hat{m}^*-\gamma^0\hat{\mu}^*   
\end{array}   
\right],   
\label{eigmatrix}   
\end{equation}   
in color, flavor, Dirac, and Nambu-Gorkov space. Here, $\Delta_{AA}$   
are the diquark gaps. $\hat{m}^*$ is the diagonal renormalized mass   
matrix and $\hat{\mu}^*$ the renormalized chemical potential matrix,   
$\hat{\mu}^*={\rm diag}_f  
(\mu_u-G_S\eta_V\omega_0,\mu_d-G_S\eta_V\omega_0,\mu_s-G_S\eta_V\phi_0)$.   
The gaps and the renormalized masses are determined by minimization   
of the mean-field thermodynamic potential \eqn{potential}.   
We have to obey constraints of charge neutrality which depend on the   
application we consider.   
In the (approximately) isospin symmetric situation of a heavy-ion collision,  
the color charges are neutralized, while the electric charge in general is   
non-zero.   
For matter in $\beta$-equilibrium, also the   
electric charge is neutralized. For further details, see   
\cite{Blaschke:2005uj,Ruster:2005jc,Abuki:2005ms}.   
    
We consider $\eta_D$ as a free parameter of the quark matter   
model, to be tuned with the present phenomenological constraints on   
the high-density EoS.   
Similarly, the relation between the coupling in the scalar and   
vector meson channels, $\eta_V$, is considered as a free parameter   
of the model.   
The remaining degrees of freedom are fixed according to the NJL model   
parameterization in Table I of \cite{Grigorian:2006qe}, where a fit to   
low-energy phenomenological results has been made.   
   
A consistent relativistic approach to the quark hadron phase transition   
where the hadrons appear as bound states of quarks is not yet developed.   
First steps in the direction of such a unified approach to quark-hadron matter  
have been accomplished within the NJL model in \cite{Lawley:2006ps}.  
In this paper, however, the role of quark exchange interactions between   
hadrons (Pauli principle on the quark level) has yet been disregarded.  
As has been demonstrated within a nonrelativistic potential model approach,  
these contributions may be essential for a proper understanding of the   
short-range repulsion \cite{Ropke:1986qs} as well as the asymmetry energy   
at high-densities \cite{Blaschke:1989nn}.   
In the present work we apply a so-called two-phase description where   
the nuclear matter phase is described   
within the relativistic Dirac-Brueckner-Hartree-Fock (DBHF) approach   
considered in \cite{Klahn:2006ir} and  
the transition to the quark matter phase given above is obtained by   
a Maxwell construction.  
The critical chemical potential of the phase transition is obtained from  
the equality of hadronic and quark matter pressures.  
A discussion of the reliability of the Maxwell construction for the case of   
a set of conserved charges is discussed in \cite{Voskresensky:2002hu}.  
   
\begin{figure}[thb]      
\includegraphics[width=0.7\figurewidth,angle=-90]{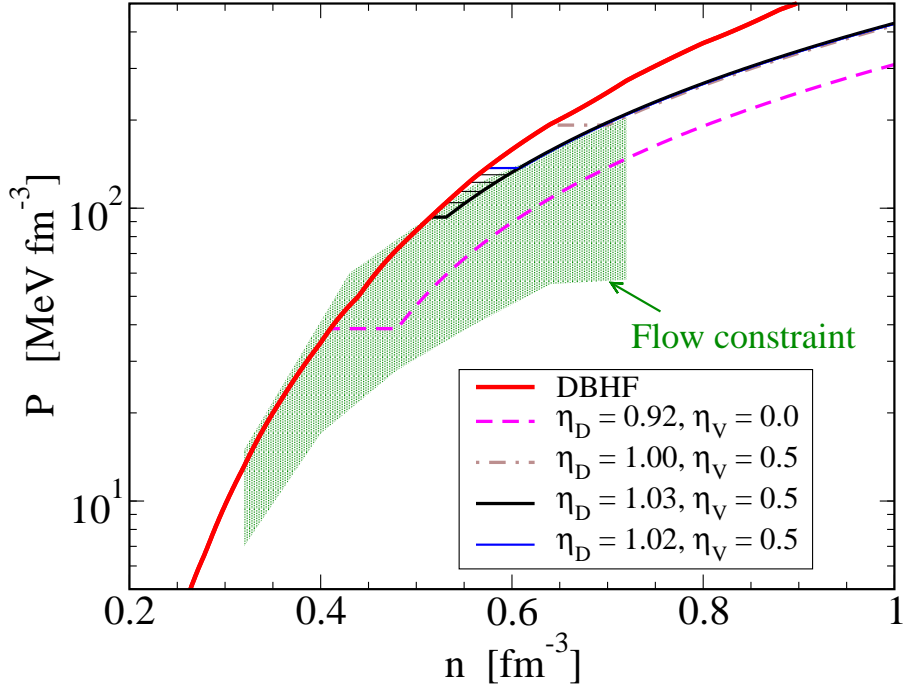}   
\caption{Pressure vs. density of the isospin symmetric EoS for different   
values of the relative coupling strengths $\eta_D$ and $\eta_V$.   
The behaviour of elliptic flow in heavy-ion collisions is related   
to the EoS of isospin symmetric matter. The upper and lower limits   
for the stiffness deduced from such analyses are indicated in the   
figure (shaded region).  
The quark matter EoS favored by the flow constraint has a vector coupling   
$\eta_V=0.50$ and a diquark coupling between $\eta_D=1.02$ (blue solid line)  
and $\eta_D=1.03$ (black fat solid line); results for four intermediate  
values $\eta_D=1.022~\dots~1.028$ are also shown (thin solid lines).}   
\label{fig:dan}   
\end{figure}      
The baryon density as derivative of the pressure with respect to the   
baryochemical potential exhibits a jump at the phase transition, as   
shown for isospin-symmetric matter in Fig.~\ref{fig:dan}.  
As can be seen in that Figure, a slight variation of the quark matter  
model parameters  $\eta_D$ and $\eta_V$ results in considerable changes   
of the critical density for the phase transition and the behaviour of the   
pressure (stiffness) at high densities.  
There appears the problem of a proper choice of these parameters which we   
suggest to solve by applying the flow constraint \cite{Danielewicz:2002pu}  
to symmetric matter, shown as the hatched area in Fig.\ref{fig:dan}.   
At first we fix the vector coupling by demanding that the high  
density behavior of the hybrid EoS should be as stiff as possible but still in   
accordance with the flow constraint.   
We obtain  $\eta_V=0.50$, independent of the choice of the scalar diquark   
coupling.  
The latter we want to determine such that the problem of the  violation  
of the flow constraint for the DBHF EoS in symmetric nuclear matter   
at high densities is resolved by the phase transition to quark matter.  
The optimal choice for $\eta_D$ is thus between 1.02 and 1.03.  
In the following we will investigate the compatibility of the now defined  
hybrid star equation of state with CS constraints.  
  
\section{Astrophysical constraints on the high-density EoS}     
\label{sec:constraints}     
Recently,  observations of compact objects have resulted in   
new limits for masses and radii which put stringent constraints on the   
high-density behaviour of the nuclear matter EoS, see \cite{Klahn:2006ir}.  
  
\begin{figure}[htb]      
\includegraphics[height=\figurewidth,angle=-90]{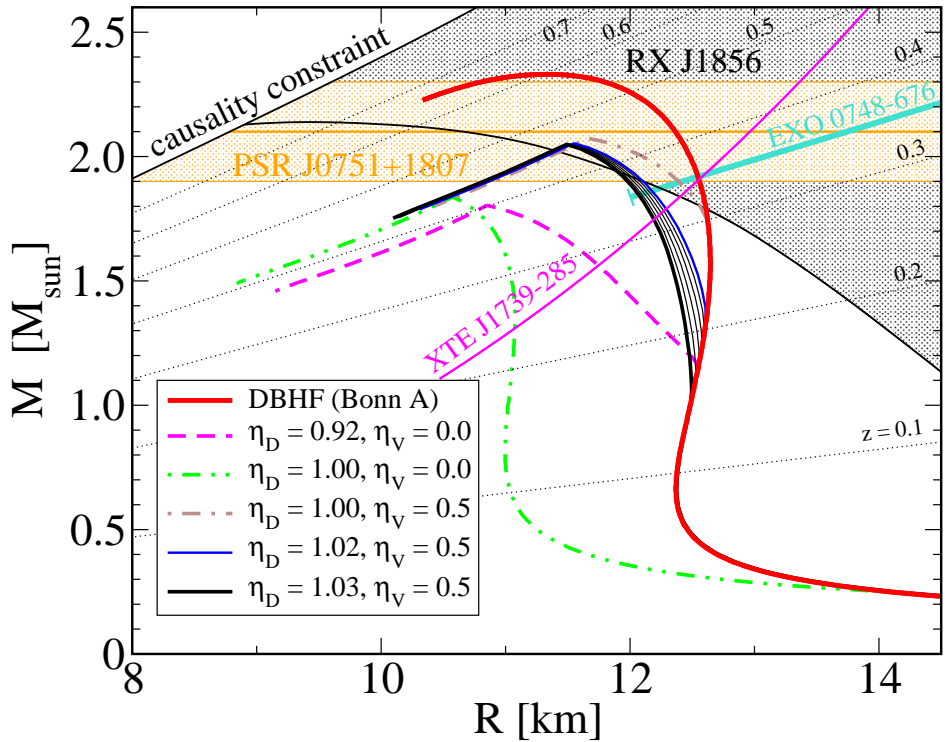}  
\caption{    
Mass - radius relationship for CS sequences corresponding   
to a nuclear matter EoS  (DBHF) and different hybrid   
star EsoS (DBHF+NJL), see text. Indicated are also the constraint on the mass   
from the pulsar J0751+1807 \protect{\cite{NiSp05}}   
and on the mass-radius relationship from   
the isolated neutron star RX J1856  \protect{\cite{Trumper:2003we}}  
Present constraints   
on the mass-radius relation of CSs do not rule out hybrid stars.   
The dotted lines indicate the gravitational redshift, $z=(1-2GM/R)^{-1/2}-1$,   
of photons emitted from the compact star surface.}   
\label{fig:m-r}    
\end{figure}

Particularly demanding data come from the pulsar PSR J0751+1807 with a lower   
mass limit of $M\ge 1.9$~M$_\odot$ \cite{NiSp05},   
and the isolated compact star RX J1856  with a mass-radius relationship   
supporting a radius exceeding 13.5 km for typical masses below $1.4~M_\odot$   
or masses above $1.9~M_\odot$ for stars with radii $R\le 12$ km   
\cite{Trumper:2003we}.   
In Fig.\ref{fig:m-r} we display these constraints together with lines of   
constant gravitational redshift as, e.g., for the putative surface redshift   
measurement \cite{Cottam:2002} of EXO 0748-676.   
  
The above constraints have to be explained by any reliable CS EoS,     
i.e. the mass radius relation resulting from a corresponding solution of the   
Tolman-Oppenheimer-Volkoff equations has to    
touch each of the regions shown in  Fig.~\ref{fig:m-r}.    
This is well fulfilled for the purely hadronic DBHF EoS.    
  
It is widely assumed that if quark matter would exist in CSs, the   
maximum mass would be significantly lower than for nuclear matter stars (NMS).  
This argumentation has been used to claim that quark matter in neutron   
stars is in contradiction with observations \cite{Ozel:2006km}.  
  
As we will show in this work, large hybrid star masses can be obtained for   
sufficiently stiff quark matter EsoS.  
In this case, the corresponding hybrid (NJL+DBHF) QCS sequence is { not}   
necessarily ruled out by phenomenology.   
The stiffness of the quark matter EoS can be significantly increased  
when the vector meson interaction   
of the NJL model introduced in Section \ref{sec:eos} is active.   
The maximum value of the vector coupling which is still in accordance with the   
upper limit extrapolation of the flow constraint, $\eta_V=0.50$, see Fig.~1,   
allows a maximum mass of  $2.1~M_\odot$.  
With this choice the constraints from  
PSR J0751+1807 and RX J1856 displayed in Fig.~2 can be fulfilled.   
  
The maximum mass is rather inert to changes of the diquark coupling  $\eta_D$  
whereas the critical mass for the occurrence of a quark matter core gets   
significantly lowered by increasing the value of $\eta_D$.  
For example, the choice of $\eta_D$ in the range $1.02~-1.03$ corresponds to   
critical star masses from $1.35~M_\odot$ to $1.0~M_\odot$, see Figs.~2 and 4.   
   
We find for all hybrid EoS studied in this paper that  
the occurrence of a CFL quark matter core renders the compact star   
unstable.  
This confirms earlier findings by Baldo et al. \cite{Baldo:2002ju} and   
Buballa et al. \cite{Buballa:2003et} for slightly different hybrid EsoS.    
  
An additional test to the mass-radius relationship is provided by a  
measurement of the gravitational redshift of line emissions from the   
CS surface.   
The disputed measurement of $z=0.35$ for    EXO 0748-676   
\cite{Cottam:2002} would be in accordance with both NMS and QCS   
interpretations. A measurement of $z\ge 0.5$ could not be accommodated with   
the QCS model suggested here, while the NMS would not be invalidated by   
redshift measurements up to $z=0.6$.  

\begin{figure}[htb]      
\includegraphics[height=\figurewidth,angle=-90]{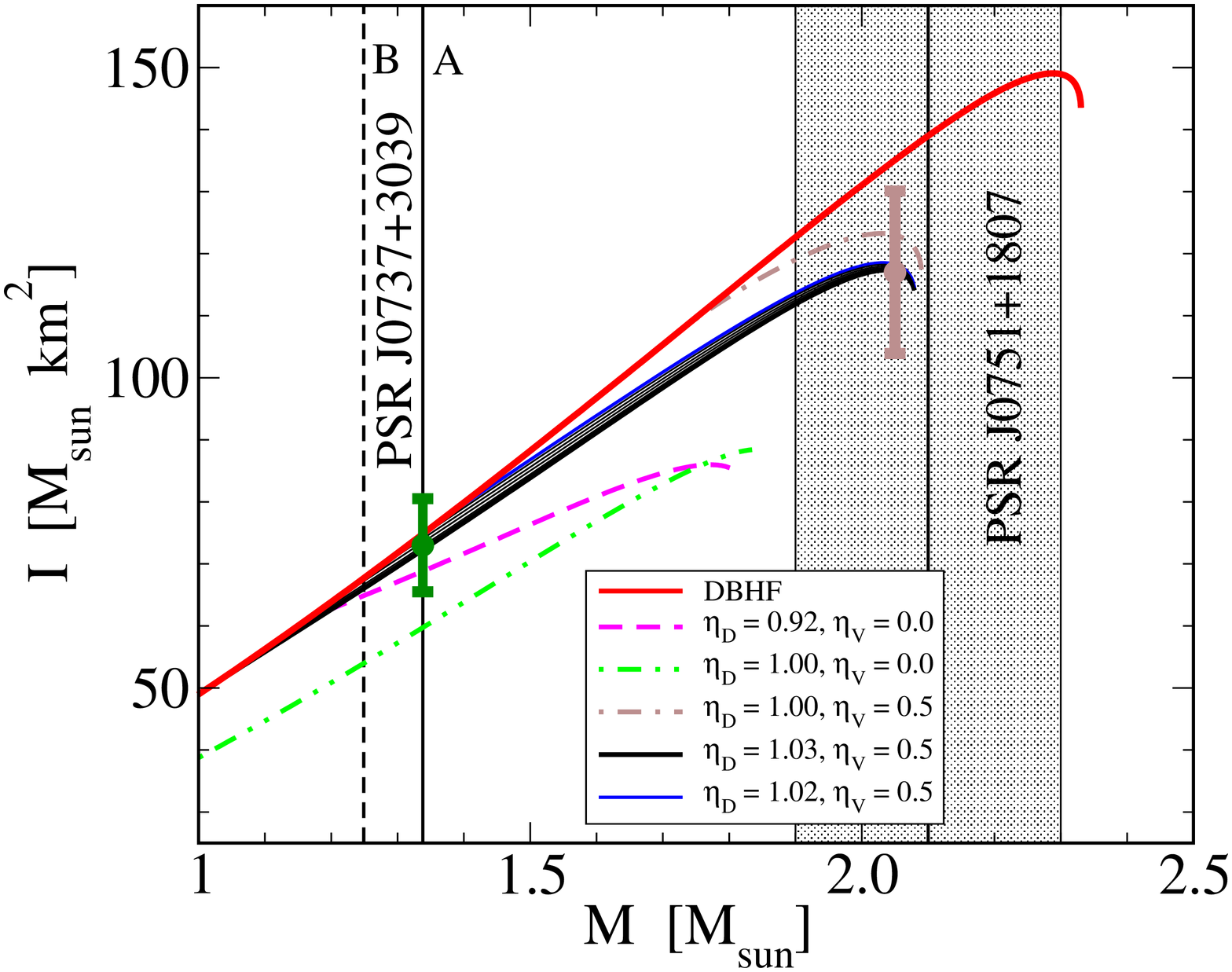}      
\caption{\label{fig:moi}     
Moment of inertia vs. the gravitational mass for the compact star   
EsoS discussed in the text.   
The highlighted mass regions correspond to the double pulsar   
J0737+3039 A+B and the pulsar J0751+1807.  
We also show anticipated data points with error bars corresponding to  
a measurement with a 1 $\sigma$ level of 10\%.  
}    
\end{figure}      
Next we want to discuss the    
question whether measurements of the moment of inertia~(MoI)~$I$    
might serve as a tool to distinguish pure NMS from QCS models.    
Due to the discovery of the relativistic double pulsar  PSR J0737+3039  
a measurement of the MoI became a possibility and has been recently discussed   
as another constraint on the EoS of compact stars, assuming that future    
measurements will exhibit an error of only about $10\%$  
\cite{Bejger:2005jy,Lattimer:2004nj}.    
In our calculations we follow the definition of the MoI given  
in Ref. ~\cite{Weber:1999qn} and show the results for the   
EsoS of the present paper in Fig.~\ref{fig:moi}.  
Due to the fact that the mass 1.338 $M_\odot$ of  PSR J0737+3039 A  
is in the vicinity of the suggested critical mass region, the quark matter   
core is small and the expected MoI of the hybrid star will be practically   
indistinguishable from that of a pure hadronic one.    
The situation would improve if the MoI could be measured for more massive   
objects because the difference in the MoI of both alternative models for   
masses as high as $2~M_\odot$ could reach the  $10\%$ accuracy level.    
    
Finally we would like to discuss the question whether there   
are observables suited to distinguish between pure neutron stars  
and those with a quark matter interior.   
As we have seen in the previous results, the hydrodynamic behavior  
of the hybrid star EoS has to be rather similar to that of pure   
hadronic matter in order to allow for sufficiently large masses.  
If so, the moment of inertia and other mechanical properties of   
the resulting stars will turn out to be indistinguishable to the   
level of a few percent.  
A different situation might occur for CS cooling where the transport  
properties and thus the excitation spectra of the dense matter play  
the essential role. As an example, pairing gaps of the order of  
an MeV or below will not affect the thermodynamics but are sufficiently   
large to influence on neutrino cooling processes.  
Let us discuss the example of the direct Urca (DU) process.  
  
If the DU process would occur in hadronic matter, it would give rise to a   
fast cooling and result in a strong sensitivity to slightest mass   
changes of the corresponding compact object.   
Therefore, it should not occur in CSs with masses below   
$1.5$~M$_\odot$, as this would provide a cooling rate that   
is inconsistent with CS population syntheses \cite{Popov:2004ey,Popov:2005xa}.   
If on the other hand the DU process does {\it not} occur in a hadronic star,  
one would require that young, fast coolers such as Vela and 3C58 should have   
a rather large mass, again in contradiction with the present population   
syntheses  
\cite{Blaschke:2006gd}.  
  
A possible resolution to this hadronic cooling problem  
could be a phase transition to quark matter with moderately enhanced cooling.  
This has been demonstrated for hybrid stars with a 2SC+X quark matter phase   
 \cite{Grigorian:2004jq}  
which is in accordance with all presently known cooling constraints   
\cite{Popov:2005xa,Blaschke:2006gd}.   
The physical nature of the hypothetical X-gap is, however, not yet clarified.   
A discussion of this issue can be found in Refs.  
\cite{Blaschke:2005dc,Blaschke:2006tt}.  
   
\begin{figure}[htb]      
\includegraphics[width=0.7\figurewidth,angle=-90]{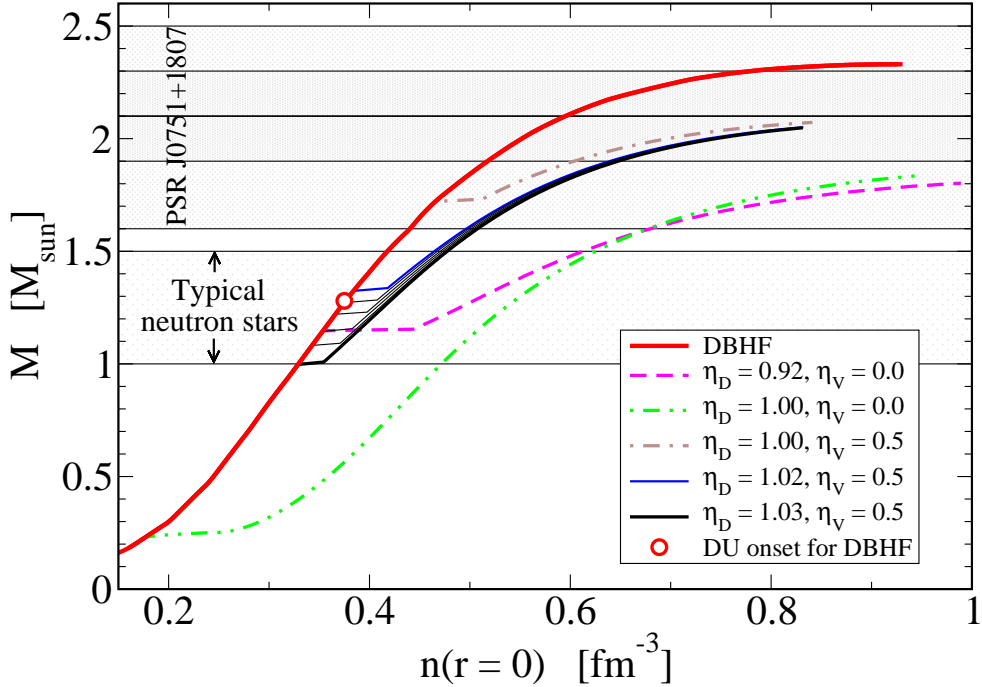}      
\caption{    
Compact star masses vs. the central density for different   
values of the relative coupling strengths $\eta_D$ and $\eta_V$,   
see Section \ref{sec:eos}. In a pure neutron (DBHF) star,      
DU processes occur for $M>M_{\rm DU}=1.26$~M$_\odot$ (circle).   
The maximum mass of hybrid star configurations depends mainly   
on the relative strength of the vector meson coupling, $\eta_V$,    
while the transition density depends on $\eta_V$ and $\eta_D$.}    
\label{fig:m_n}   
\end{figure}

For the DBHF EoS the DU threshold is at  
$n_{\rm DU} = 0.375$~fm$^{-3}$ corresponding to a CS mass of   
$M_{\rm DU} = 1.26$~M$_\odot$, see Fig.~\ref{fig:m_n}.   
The hybrid EoS presented in this work has a critical density   
for the quark matter phase transition which is below $n_{\rm DU}$  
provided  a value $\eta_D \ge 1.024$ is chosen.  
  
Thus for the parameter values $\eta_V =0.50$ and $\eta_D \gsim 1.024$   
the present EoS for hybrid star matter fulfills all   
modern observational constraints discussed above.  
  
\section{Summary}    
We have investigated the compatibility of present constraints    
on the high density EoS with the concept of CSs possessing a QM core.    
The hadronic part of the EoS was taken from  the DBHF approach,    
while the QM EoS is provided by a chiral NJL-type quark model    
with current-current interactions     
in the color singlet isoscalar scalar and vector    
meson channels as well as in the color antitriplet scalar diquark      
channels.    
The finite vector meson coupling  enables us to describe large hybrid star   
masses by stiffening the QM EoS,    
whereas the chosen value for the diquark coupling    
ensures a sufficiently early phase transition to QM in order    
to avoid the activation of DU cooling within the hadronic shell of the   
hybrid star.    
Since the high density QM part of the EoS is softer     
than the corresponding pure hadronic EoS     
also the flow constraint is fulfilled in this scenario.    
We discussed the possibility to distinguish between    
pure NSs and hybrid stars by hypothetical data from successful measurements    
of the MoI for stars of a given mass.    
It turned out, that this would be possible    
for rather massive objects ($M\approx 2.0~M_\odot$)    
provided the standard deviation of the  measurements is less than $10\%$,    
as expected for PSR J0737+3039 A.    
    
As our main result we conclude that    
{\it no present phenomenological finding    
bears a strong argument against the presence of    
a QM core inside NSs.}    
Moreover, we demonstrated that problems with cooling and flow which appear   
as weak points of  a purely hadronic EoS at large densities     
can be resolved in a natural way when a transition to QM    
occurs at not too large densities.  
    
The most common argument against the presence of QM in CSs,    
resulting from the {\it prejudice} of the softness of    
QM EsoS, is no longer valid if we account for    
a vector meson interaction channel which stiffens the EoS.    
The earlier finding \cite{Baldo:2002ju,Buballa:2003qv} that the   
occurrence of CFL quark matter renders hybrid stars unstable  
has been reconfirmed and therefore it should become customary   
for phenomenological scenarios of compact star evolution based on   
the assumption of a CFL quark matter phase (see, e.g., Refs.  
\cite{Alford:2004zr,Berdermann:2005yn,Drago:2005qb,Page:2005ky})  
to check the necessary constraint of stability against gravitational collapse.  
   
The present study of dense quark matter in compact stars, albeit based  
on a microscopic dynamical approach used the strongly simplified NJL model  
interaction.   
It is desireable to repeat the present study for more realistic microscopic   
approaches to the QCD EoS, e.g., on the basis of the QCD Dyson-Schwinger   
equation approach where interactions are nonlocal and momentum-dependent.  
One intermediate step towards this goal could consist in using Lattice QCD  
results on the momentum dependence of the quark selfenergies  
\cite{Parappilly:2005ei} to adjust covariant formfactors of nonlocal,  
separable models \cite{Blaschke:2004cc,GomezDumm:2005hy}  
for the equation of state of quark matter in compact stars.  
\label{subsec:sum}     

\subsection*{Note added in proof}

After receiving the proofs for this Letter we have been informed that the mass 
of pulsar J0751*1807 has been corrected by Nice \cite{Nice:2007} using a 
longer data base which allows to improve the value of the orbital decay and 
the detection of the Shapiro delay. Instead of previously 
$2.1 \pm 0.2~M_\odot$ $(1\sigma)$ the new value is $1.26~M_\odot$ with 
estimated errors from $1.12-1.30~M_\odot$ $(1\sigma)$ and  $0.98-1.53~M_\odot$ 
$(2\sigma)$. 
Therefore the mass of this object does no longer hold as a constraint on the 
EOS as described above. However, other new observations provide mounting 
evidence for the existence of a high maximum compact star mass.
In addition to the LMXB 4U 1636-536 with   $2.0 \pm 0.1~M_\odot$ 
\cite{Barret:2005wd} mentioned in the introduction, very recent measurements 
on millisecond pulsars in globular clusters yield $M=1.96 +0.09/-0.12~M_\odot$ 
and      $2.73 \pm 0.25~M_\odot$ $(1\sigma)$ for PSR B1516+02B and 
PSR J1748-2021B, respectively \cite{Freire:2007}.
We also note that recent observations of the thermal polar cap emission of the 
quiescent LMXB X7 in the globular cluster 47 Tuc yield a rather large radius 
of $14.5+1.8/-1.6$ km for a 1.4 solar mass compact star \cite{Rybicki:2005id},
which comes close to the lower bound on the mass-radius relation provided by 
RX J1856-3754 which is quoted in our Letter.
In summary we conclude that the large radii and masses found in several 
compact stars require a rather stiff equation of state.
 
\section*{Acknowledgments}   
We are grateful for critical reading and comments by M. Alford, M. Buballa,  
A. Drago, G. Grunfeld and F. Weber.  
We acknowledge the support by the Virtual Institute of the Helmholtz   
Association under grant VH-VI-041. The work of H.G. was supported in part  
by DFG grant No. 436 ARM 17/4/05 and that of F.S. by the Swedish Graduate   
School of Space Technology and the Royal Swedish Academy of Sciences.   
D.B. is supported by the Polish Ministry of Science and Higher Education.

\end{document}